\documentclass{emulateapj}
\usepackage{multirow}
\usepackage[colorlinks,urlcolor=blue,citecolor=blue,linkcolor=blue]{hyperref}
\usepackage{graphicx}
\usepackage{amsmath}
\usepackage{threeparttable}

\begin{document}
\title{Electron acceleration in One-dimensional non-relativistic quasi-perpendicular collisionless shocks}

\author{Rui Xu}
\author{Anatoly Spitkovsky}
    \affiliation{Department of Astrophysical Sciences, Peyton Hall, Princeton University, Princeton, NJ, 08544}
\author{Damiano Caprioli}
  \affiliation{Department of Astronomy and Astrophysics, University of Chicago, 5640 S. Ellis Ave, Chicago, IL 60637}
\shortauthors{Xu et al.}

\begin{abstract}
We study diffusive shock acceleration (DSA) of electrons in non-relativistic quasi-perpendicular shocks using self-consistent one-dimensional particle-in-cell (PIC) simulations. By exploring the parameter space of sonic and Alfv\'{e}nic Mach numbers  we find that high Mach number quasi-perpendicular shocks can efficiently accelerate electrons to  power-law downstream spectra with slopes consistent with DSA prediction. Electrons are reflected by magnetic mirroring at the shock and drive non-resonant waves in the upstream.  Reflected electrons  are  trapped  between  the  shock  front  and  upstream waves, and undergo multiple cycles of shock drift acceleration before the injection into DSA. Strong current-driven waves also temporarily change the shock obliquity and cause mild proton pre-acceleration even in quasi-perpendicular shocks, which otherwise do not accelerate protons. These results can be used to understand nonthermal emission in supernova remnants and intracluster medium in galaxy clusters. 
\end{abstract}


\maketitle

\section{Introduction}

Non-thermal particles are ubiquitous in the Universe. The acceleration of these particles is often associated with collisionless shocks.  For example, it is widely regarded that supernova remnant (SNR) shocks are responsible for the acceleration of galactic cosmic rays (CRs) up to the knee $\rm E\sim 10^{16} eV$ (e.g., \citealp{gaisser2016}).
Evidence of electron acceleration in collisionless shocks has also been provided by numerous in-situ observations of high Mach number shocks in the heliosphere (e.g., at Saturn's bow shock
by the Cassini spacecraft, \citealp{sulaiman2015,masters2017}). 
 The dominant acceleration mechanism in astrophysical shocks is thought to be due to the diffusive shock acceleration process (DSA, \citealp{Bell1978, BlandfordOstriker1978,Drury1983,BlandfordEichler1987}), where particles gain energy by repeatedly crossing the shock while scattering off converging magnetic perturbations on both sides. The final momentum distribution $f(p)$ is a universal power-law with index $f(p) \propto p^{-3r/(r-1)}$,  where $r$ is the shock compression ratio. For strong shocks with $r=4$, the momentum distribution follows $f(p)\propto p^{-4}$.

While DSA naturally produces power-law distributions, it works only for particles whose Larmor radius is larger than the shock transition width, which is typically of the order of proton gyro-radius. One of the most important questions in CR physics, known as ``the injection problem," is how particles are extracted from the thermal pool to participate in DSA.  Achieving injection energy is more challenging for electrons due to their smaller Larmor radii, compared to protons. Also, the shock potential barrier is tuned to reflect upstream ions, which hinders electron reflection \citep{CaprioliPop2015}.

Proton and electron injection for non-relativistic {\it quasi-parallel} shocks,  where  the angle between the background magnetic field and the shock normal is $\theta<45^\circ$, have been studied with fully kinetic PIC simulations that show both species successfully injected into DSA \citep{Kato2015, ParkSpitkovsky2015}. 
For {\it quasi-perpendicular} shocks ($\theta>45^\circ$), proton acceleration and reflection into the upstream has been shown to be inefficient without pre-existing upstream turbulence (\citealp{CaprioliSpitkovsky2014,CaprioliPop2015,CaprioliZhang2018}). In quasi-perpendicular shocks with low sonic Mach numbers, electron scattering was reported to be mediated by oblique electron firehose instability driven by electrons reflected from the shock \citep{GuoSironi2014a, GuoSironi2014b}. Electron pre-acceleration was also observed in perpendicular and quasi-perpendicular multi-dimensional PIC simulations \citep{MatsumotoAmano2017,BohdanNiemiec2017}. However, previous studies have not shown compelling evidence of DSA spectra forming downstream of quasi-perpendicular shocks.

In this Letter, we study the formation of DSA power-law $f(p)\propto p^{-4}$ for electrons downstream of quasi-perpendicular collisionless shocks using 1D PIC simulations. The simulations run long enough to demonstrate successful electron injection into DSA in the absence of substantial proton acceleration. We show how electrons are extracted from the thermal pool and injected into DSA by scattering on electron-driven waves in the upstream. Finally, we discuss the effect of sonic and Alfv\'{e}nic Mach numbers of shocks on the downstream electron spectra. 

\section{Simulation setup}

We performed numerical simulations with the electromagnetic PIC code TRISTAN-MP \citep{Spitkovsky2005}. To enable long integration times, the simulation domain is 1D along $\hat{x}$ direction, retaining all components of fields and velocities. The setup is very similar to previous PIC simulations of collisionless shocks (e.g., \citealp{Spitkovsky2008a,SironiSpitkovsky2011,ParkSpitkovsky2015}). In order to facilitate the analysis of upstream waves, the simulations are performed in the upstream rest frame by moving the left conducting boundary wall into a stationary plasma. The resolution of our reference run is 10 cells per electron skin depth, $c/\omega_{pe}$, where $c$ is the speed of light and $\omega_{pe} = \sqrt{4\pi n e^2/m_e}$ is the electron plasma frequency ($e$, $n$ and $m_e$ are electron charge, number density and mass, respectively). The simulation domain is enlarged by expanding the right boundary with time to save computational resources, with the final domain size reaching $\sim 3.5\times10^4 c/\omega_{pe}$. We use 256 particles per cell per species with a reduced proton-to-electron mass ratio $m_p/m_e=100$. The wall velocity is fixed at $v_0=0.15c$. Upstream protons and electrons are assumed to be in thermal equilibrium with temperature $T_p = T_e = 4\times 10^{-4}m_e c^2$. The corresponding sonic Mach number is $M_s = v_{sh}/\sqrt{\gamma (T_p+T_e)/m_p} \approx 55$ for adiabatic index $\gamma=5/3$. The Alfv\'{e}nic  Mach number is $M_A = v_{sh}/v_A \approx 63$, where $v_A = { B_0}/\sqrt{4\pi n_0 m_p}$ is the Alfv\'{e}n speed for the initial magnetic field, ${\bf B_0} = B_0(\cos\theta\hat{\bf x} + \sin\theta\hat{\bf y})$, inclined at an angle $\theta = 63^\circ$ relative to the shock normal, so the shock is quasi-perpendicular.

\begin{figure}
    \centering
    \includegraphics[width=0.5\textwidth]{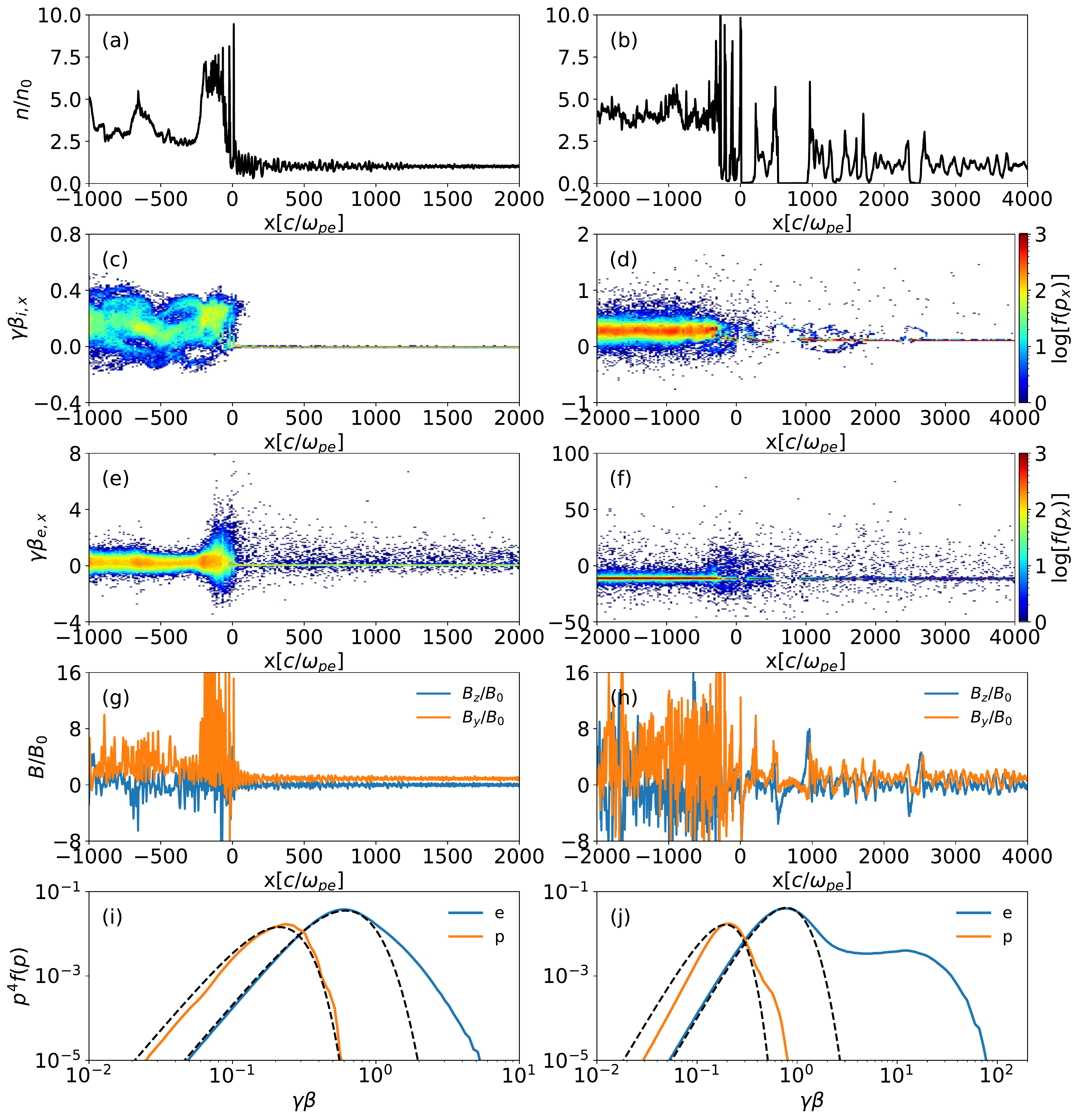}
    \caption{Structure of quasi-perpendicular shock ($\theta = 63^\circ$, $M_s$ and $M_A$ $\sim60$) at simulation time $\sim 10\Omega_{cp}^{-1}$ (left panels) 
    and $\sim 45\Omega_{cp}^{-1}$ (right panels): (a-b) electron number density 
normalized by the upstream value;  (c-f) proton and electron  $x-p_x$ phase space distribution 
$f(p_x)$; (g-h) $\hat{y}$ and $\hat{z}$ components of the magnetic field normalized by the background magnetic field; (i-j) downstream electron and proton spectra, where the dashed lines represent thermal Maxwellian distributions.}
    \label{fig:phase}
\end{figure}

\section{Results}

Figure \ref{fig:phase} shows the shock structure at time $t\approx 3.2\times10^4\omega_{pe}^{-1}\approx 10\Omega_{cp}^{-1}$ (left panels) and at the final time $t\approx 45\Omega_{cp}^{-1}$ (right panels), where $\Omega_{cp}=e B_0/m_p c$ is the proton-cyclotron frequency. Electron number density (Fig.~\ref{fig:phase}a,b) is compressed by a factor of $r\approx 4$ in the far downstream region, as expected. The density overshoot at the shock is attributed to gyrating protons undergoing coherent motion 
 (e.g., \citealp{LeroyGoodrich1981, Wu1984}). In Fig.~\ref{fig:phase}c-f, we show the proton and electron $x-p_x$ phase space distribution. Protons are reflected by the potential barrier at the shock, but due to magnetic obliquity they cannot escape far upstream (Fig.~\ref{fig:phase}c). Unlike protons, the reflected electrons can outrun the shock. The energy gain during the magnetic mirroring at the shock ramp region increases the projected velocity along the shock normal, which can be larger than the shock propagation speed \citep{BallMelrose2001, ParkRen2013}.
The reflected electrons preferentially move along the background magnetic field and contribute a net flux along the magnetic field lines. The reflected electrons drive strong waves in the upstream field which can scatter electrons back to the shock (see Fig.~\ref{fig:phase}g,h). Figures \ref{fig:phase}(i-j) show the downstream electron and proton spectra. We see electrons successfully injected into DSA while protons only form a steep non-thermal tail.

\begin{figure}
    \centering
    \includegraphics[width=0.4\textwidth]{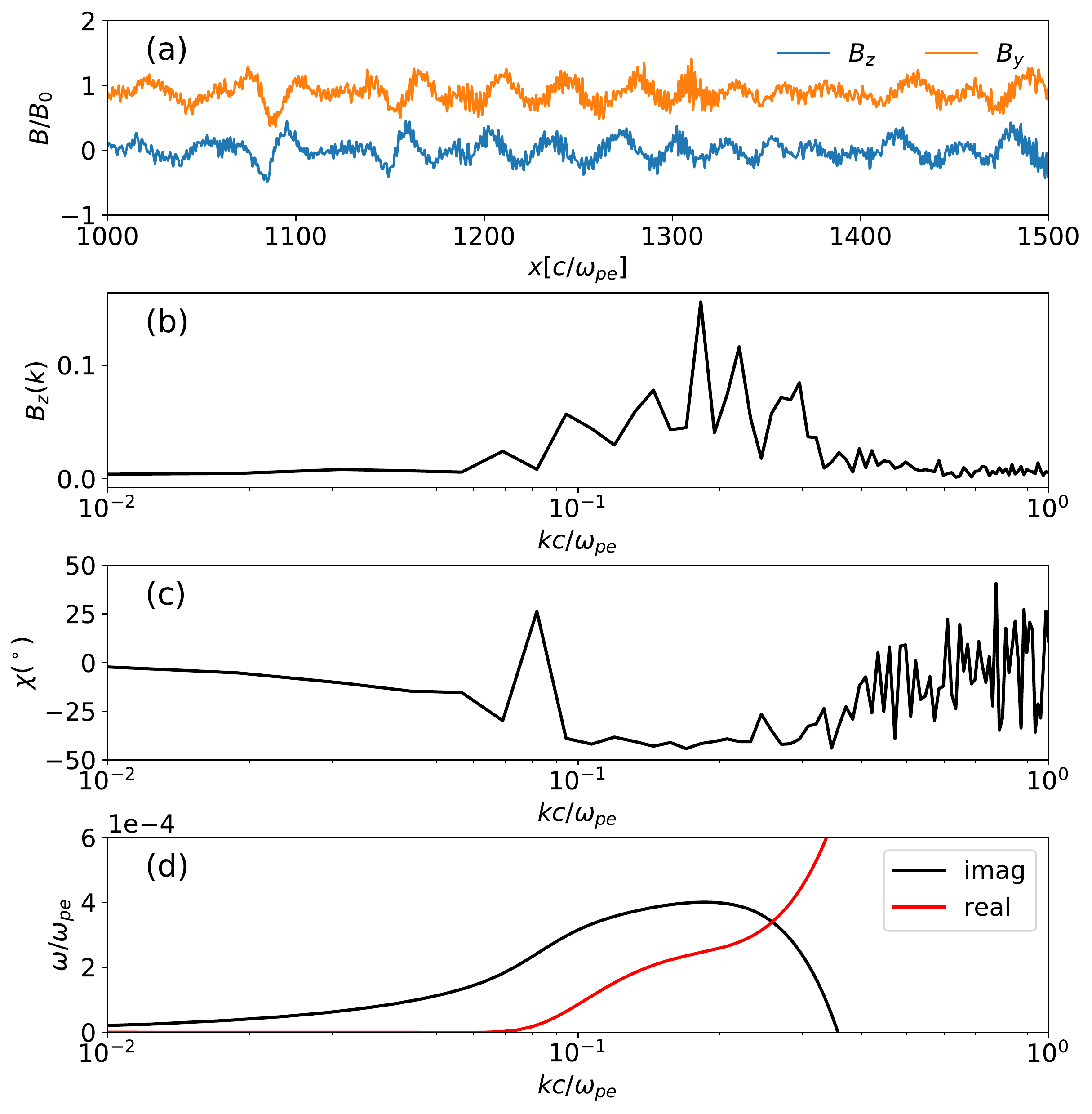}
    \caption{(a) Normalized magnetic field in the upstream region, with $x$ coordinate measured relative to the shock ramp. (b) Fourier transform of $B_z$ in the 
    same region. (c) Polarization angle $\chi$ of the upstream wave, where $\chi = \pm 45^\circ$ corresponds to right-(left-)handed circularly polarized modes. The wave is left hand circularly polarized and, thus, non-resonant with electrons. (d) Real (red line) and imaginary (black line) part of the plasma
dispersion relation for the beam-plasma system, where the number fraction of the beam is $n_b/n_0=0.03$ and the beam drift velocity is $v_{dr} = 0.35c$. The wavelength at the maximum growth rate roughly matches the wavelength from shock simulations. }
    \label{fig:wave}
\end{figure}

To study the nature of the electron-driven waves, we perform Fourier analysis of the z-component of magnetic field in the upstream region, as shown in Fig.~\ref{fig:wave}b. We see that the spectral energy density peaks at $k\simeq 0.1-0.2\omega_{pe}/c$. The reflected electrons are magnetized with mean gyroradius smaller than the wavelength. Figure \ref{fig:wave}c shows the wave polarization angle  $\chi= 0.5\times \sin^{-1}(\bf V/I)$, where $\bf I$, $\bf V$ are the Stokes parameters for the two transverse magnetic field components. The angle $\chi = \pm 45^\circ$ corresponds to a right-(left-)handed circularly polarized wave. We see that the wave is left-hand circularly polarized and, thus, non-resonant with electrons.  The instability responsible for driving upstream waves is very similar to the electron heat flux instability studied in solar wind physics (e.g., \citealp{Gary1975, saeed2016, lee2019}). Electron heat flux can excite right-hand polarized whistler waves and left-hand polarized firehose waves depending on the speed of heat-carrying ``beam" electrons \citep{gary1985, shaaban2018}. Larger velocities of beam electrons make them less resonant, inhibiting the whistler heat flux mode and exciting the firehose mode.

\cite{GuoSironi2014b} studied firehose-mediated electron acceleration in the low Mach number high-beta plasma. While the difference between modeling the system with a single bipolar distributed plasma and a thermal background with an electron beam is not significant for low Mach number high-beta plasma, the two become more distinct in shocks with higher Mach number and low plasma beta, where the beam appears more separated in phase space. Following \cite{Gary1975} and \cite{stix1992}, we computed the kinetic plasma dispersion relation for a three-component plasma (beam electrons, background electrons moving in the opposite direction to enforce current neutrality, and background protons).  Similar analysis has been done for electron firehose instability in high-beta intracluster shocks \citep{Kim2020} and counter streaming electron beams \citep{lopez2020firehose}.
Calling $n_b$ and $n_0$ the beam and background number densities, we fix the number fraction of the beam $n_b/n_0\simeq 0.03$ and the average beam velocity along the magnetic field $v_{dr} \simeq 0.35c$, as measured for the electron beam in our benchmark simulation. The beam electrons are modeled as drifting Maxwellian with temperature $100$ times the background electron temperature. We compute the waves propagating obliquely to the background magnetic field at angle $\theta=63^\circ$,
 which are the waves that can propagate along the x-axis of our 1D simulations. The firehose-type instability growth rate from the dispersion relation is shown in Figure \ref{fig:wave}d.
The linear analysis returns the fastest-growing wave with wavelength $k\simeq 0.1-0.2\omega_{pe}/c$, which is roughly consistent with the peak of Fourier spectrum of waves in the simulation (Fig.~\ref{fig:wave}a).  While a small fraction of reflected protons is observed in the shock upstream at later times, the current carried by protons is less than ten percent of the electron current and does not affect the wave nature significantly. Thus, we conclude that the waves in the upstream are firehose waves driven by returning electrons via the heat flux instability.

\begin{figure}
    \centering
    \includegraphics[width=0.45\textwidth]{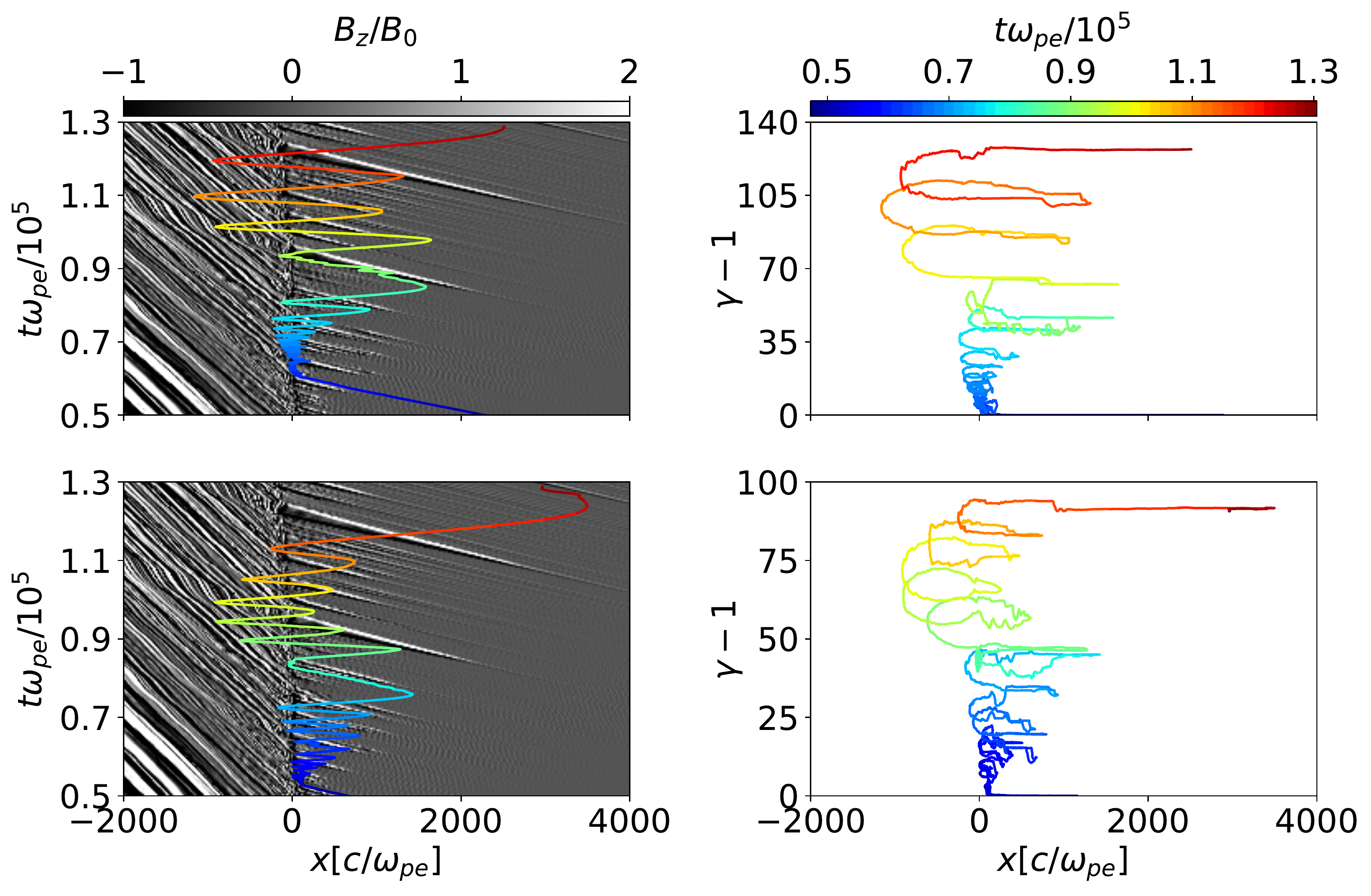}
    \caption{Electron trajectories in the space-time (left panels) and the space-energy (right panels) plots. Both electrons are injected into DSA after multiply cycles of SDA. The gray-scale color map indicates normalized z-component of the magnetic field. The color line indicates time, as in the legend.}
    \label{fig:prtl}
\end{figure}

In order to illustrate how electrons are injected from the thermal pool, we track individual particles along their trajectories in real space and momentum space. Figure \ref{fig:prtl} shows the space-time and space-energy trajectories of two typical electrons that are injected into DSA, with space-time trajectories overplotted on top of the strength of magnetic component $B_z$, shown in grayscale.  The regular pattern in the downstream is due to advected magnetic field compressions from periodic shock reformations. Electrons are preheated in the shock foot and reflected off the shock ramp due to magnetic mirroring at time $t\approx 0.5-0.6\times 10^5\omega_{pe}^{-1}$. The preheating effect has been attributed to the interaction with Buneman waves at the shock leading edge via shock-surfing acceleration (\citealp{Amano2009,Matsumoto2012,BohdanNiemiec2017,katou2019theory}). Between $t\sim 0.6-0.9\times 10^5\omega_{pe}^{-1}$ particles remain trapped between the shock front and the upstream waves, generated by escaping electrons, and repeatedly undergo cycles of shock-drift acceleration (SDA) at the shock.
The interplay between SDA and upstream wave scattering continues to accelerate electrons and transitions to standard DSA when electron momentum reaches $p_{inj} \approx 30-80 m_e c$. The transition occurs when electrons start to diffuse in the upstream/downstream and the energy gain is from the interaction with upstream/downstream waves instead of the interaction with the shock ramp. The acceleration process is very similar to electron injection in non-relativistic quasi-parallel shocks \citep{ParkSpitkovsky2015}; the major difference is that electrons are scattered by the non-resonant waves driven by returning electrons instead of returning protons. 

\begin{figure*}[!ht]
    \centering
    \includegraphics[width=0.9\textwidth]{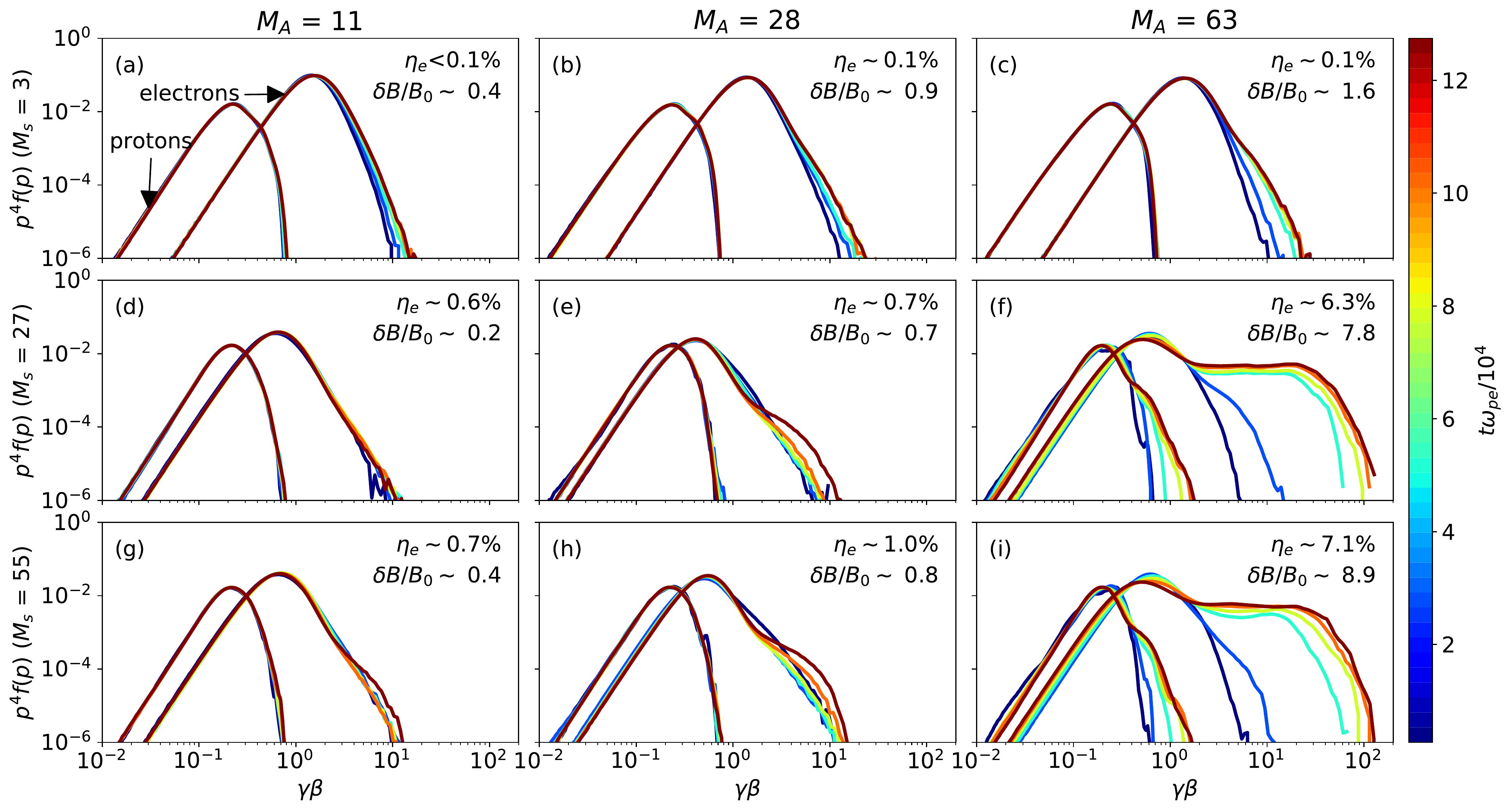}
    \caption{Downstream electron and proton spectra as a function of time for different $M_s$ and $M_A$ for quasi-perpendicular shocks  with angle $\theta=63^\circ$ and $m_i/m_e=100$. The spectrum is multiplied by $p^4$ to emphasize the scaling law expected in DSA. The color lines indicate time, as in the legend. The number fraction of non-thermal electrons $\eta_e$ at the end of the simulations, and the level of upstream magnetic fluctuations $\delta B/B_0$ are shown at the top right corner of each panel. Only shocks with $M\gg 1$ are able to produce large amplitude fluctuations with $\delta B/B_0>1$, and in these cases electrons are injected into DSA after multiple cycles of SDA and scattering of upstream waves. }
    \label{fig:comp_spect}
\end{figure*}

Figure \ref{fig:comp_spect} shows the downstream electron and proton spectra  for different Mach numbers as a function of time. The downstream electron spectrum is averaged between $200c/\omega_{pe}$ and $4000 c/\omega_{pe}$ behind the shock ramp and is multiplied by $p^4$ to emphasize the expected DSA scaling. Fig.~\ref{fig:comp_spect}i shows the downstream spectrum from our reference run, where electrons develop a power-law tail with spectral index $4$, consistent with DSA prediction (cf.~Fig.~\ref{fig:phase}j). 
In this case, electron acceleration is very efficient:  the downstream number fraction of non-thermal electrons\footnote{ We define non-thermal electrons as electrons with energy larger than five times the energy of the downstream thermal peak.} is $\eta_e \sim 7\%$, with energy fraction  $\varepsilon_e \sim 20\%$ by the end of the simulation.  Also, the maximum electron energy grows roughly linearly with time and eventually exceeds the maximum energy of downstream thermal ions. Similar to acceleration in quasi-parallel shocks, electrons show a typical DSA spectrum even in the range of momenta where they undergo SDA, indicating that the balance between energy gain and escape probability per cycle is more similar to DSA than to SDA \citep{ParkSpitkovsky2015}. Downstream protons are mostly thermal with a steep non-thermal tail, which is caused by the strong  electron-driven upstream waves that temporarily change the shock obliquity and allow a small fraction of protons to escape into the upstream. These protons are eventually advected downstream and form a steep spectrum.

Figures \ref{fig:comp_spect}a-h show the same spectra as Fig.~\ref{fig:comp_spect}i but for different $M_s$ and $M_A$. In these simulations, the left wall velocity is fixed at $v_0=0.15c$, and we change $M_s$ by varying the plasma temperature and  $M_A$ by varying the background magnetic field strength. All other parameters are as in our reference run. We see that electrons are injected into DSA in high Mach number quasi-perpendicular shocks (with both high $M_s$ and $M_A$). For shocks with low $M_s$ or $M_A$, reflected electrons also gain energy via SDA and contribute a similar upstream current but do not enter DSA at the end of the simulation.  
The acceleration efficiency depends on whether the reflected electrons are able to drive waves of large enough amplitude in the upstream.  The amplitudes of upstream magnetic fluctuations near shocks for different Mach numbers are shown in the top right corner of each panel in Fig.~\ref{fig:comp_spect}. Large Mach number and especially large $M_A$ shocks are able to drive strong upstream magnetic fluctuations with $\delta B/B_0> 1$, which are responsible for the scattering of reflected electrons (here, $\delta B$ is calculated based on the peak value of magnetic fluctuations in the upstream). For low $M_A$ shocks in hot plasma (e.g., Fig.~\ref{fig:comp_spect}a), downstream electron spectrum develops a non-thermal SDA tail. Due to the small amplitude of upstream waves, we do not observe electron injection into DSA at the end of the simulation. The threshold of the electron firehose heat flux instability follows $v_{dr}/v_A = \lambda \sqrt{m_i/m_e} \beta^{\kappa}$, where $\beta$ is the background plasma beta, and $\lambda>0$, $0<\kappa<0.1$ are constant fitting parameters \citep{shaaban2018}.  We see that it is easier to trigger the instability at lower $v_A$ and lower plasma beta (i.e., lower temperature). Hence, the heat flux instability favors large $M_s$ and $M_A$ shocks.  Indeed, we see weak waves in low Mach number shocks, and the amplitude of the waves increases with both $M_A$ and $M_s$, but depends stronger on $M_A$. The onset of the instability also depends on the ion/electron mass ratio. For realistic mass ratio, the extrapolated $M_A$ needed for electron injection into DSA may be even higher, consistent with Mach numbers of several hundred expected in SNR shocks.

In addition to Mach numbers, the efficiency of electron acceleration also depends on the shock obliquity, and is much higher in quasi-perpendicular shocks compared to quasi-parallel shocks. In the quasi-perpendicular shocks presented in this Letter,  $\eta_e \sim 0.7\%$ for Mach number $\sim30$ (see Figure \ref{fig:comp_spect}e), while for quasi-parallel shocks most of the energy goes into accelerated ions and $\eta_e\lesssim 0.1\%$ for Mach number $\sim 20$  (see Fig.~4 in \citealp{ParkSpitkovsky2015}). Such higher acceleration efficiency can be attributed to the magnetic mirror effect: electrons are more effectively reflected to the upstream in quasi-perpendicular shocks compared to quasi-parallel shocks, where the mirroring of electrons is mediated by the upstream turbulence driven by reflected protons. Also, at higher magnetic inclinations electron DSA has to vanish when obliquity approaches superluminal shocks because fewer particles can outrun the shock; we expect the optimal obliqueness angles for electron acceleration to lie between $60^\circ$ and $70^\circ$. The critical superluminal angle becomes smaller as shocks become relativistic, $v_{sh}\to c$.

\section{SUMMARY AND DISCUSSION}

We can use these findings to interpret the morphology of nontermal emission from supernova remnants. Some SNRs (e.g., SN1006) show bilateral symmetry in their synchrotron emission, which is understood as being due to pre-existing large-scale magnetic field in the remnant \citep[e.g.,][]{Reynoso2013}. 
In regions where the shock is quasi-parallel, ions are effectively injected into DSA and drive prominent magnetic field amplification with $\delta B/B_0\gg 1$, but $\eta_e$ may be $\lesssim 0.1\%$ \citep{Crumley2019,ParkSpitkovsky2015}; in these quasi-parallel regions electrons can be accelerated to multi-TeV energies, which results in non-thermal X-ray emission. In regions where the shock is quasi-perpendicular, instead, ions are not injected and the magnetic field is not effectively amplified; electron acceleration can still occur with $\eta_e\sim 5-7\%$, as in our calculations, but up to smaller energies than in quasi-parallel regions, since $\delta B/B_0\gtrsim 1$. This is consistent with the fact that quasi-perpendicular regions are radio-bright but not X-ray bright \citep{Caprioli2015, Vlasov2016}. Overall, the relative radio brightness between parallel and perpendicular regions is determined by both $\eta_e$ and $\delta B/B_0$.

An important result of this study is the existence of shocks that preferentially accelerate electrons and not ions. This helps to reduce the tension of non-detection of hadronic gamma-ray emission in galaxy clusters:
the electrons responsible for the observed radio emission could be efficiently accelerated in high $M_A$ quasi-perpendicular shocks, which do not efficiently accelerate protons. This would suppress secondary hadronic gamma-ray production. Upstream fluctuations are weaker for high $M_A$, low $M_s$ shocks, which are commonly found in clusters of galaxies \citep{MarkevVikhl2007}. According to our simulations, such shocks accelerate electrons at a lower rate, with steep power-laws. However, these shocks reflect electrons into the upstream at nearly the same rate as high $M_s$ shocks. Thus, we expect that even these weaker upstream waves may be sufficient to eventually inject electrons into DSA on substantially longer timescales. This could explain the observations where relativistic electrons produce radio ``relics" in galaxy clusters, in structures that have inferred quasi-perpendicular magnetic geometry \citep{WeerenRelic2010, Ackermann_etal_2014, BrunettiJones2014}. 

\begin{figure}[!ht]
    \centering
    \includegraphics[width=0.5\textwidth]{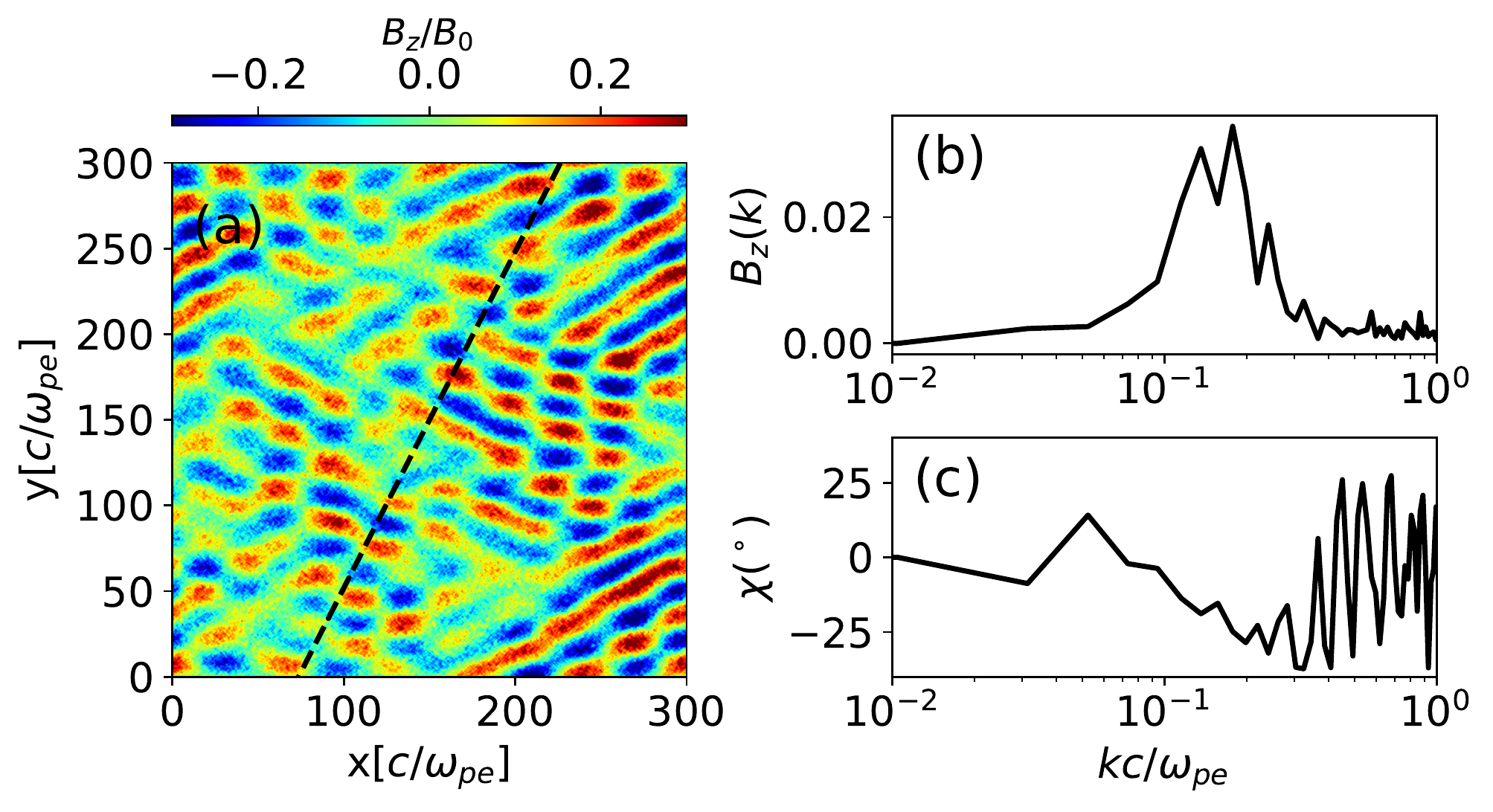}
    \caption{ 2D periodic PIC simulation for the beam plasma system with the background magnetic field along $\hat{x}$ direction. (a) normalized magnetic field $B_z/B_0$ at time $T\sim 1.3\times 10^4 \omega_{pe}^{-1}$ shows the dominant mode is oblique to the background magnetic field; (b) Fourier transform of z component of magnetic field along the direction $63^\circ$ relative to $\hat{x}$ direction (black dashed line in a); (c) polarization of the magnetic field along the same direction as panel (b).
    }
    \label{fig:wave_peri}
\end{figure}

  Although 1D PIC simulations allow us to study the long-term evolution of quasi-perpendicular shocks and to see eventual DSA spectrum formation, the applicability of 1D simulations to real systems needs to be justified. For example, 
  1D simulations limit the kinds of wave modes that can be captured when background field is inclined to the $x$ axis. This, however, is largely remedied by the fact that 
 the dominant wave mode driven by the returning electrons is oblique in nature.  To demonstrate this we performed a 2D periodic PIC simulation of an electron beam propagating along the magnetic field in the plane of the simulation (Figure \ref{fig:wave_peri}).
We initialize a static background and beam plasma with the same parameters as in our linear analysis. The current is compensated by moving the background electrons in the opposite direction.  Left panel in Figure \ref{fig:wave_peri} shows the normalized $B_z$ component at time $T\sim 1.3\times 10^4\omega_{pe}^{-1}$. We see that the dominant waves are indeed oblique to the background magnetic field that is pointing along $\hat{x}$. Right panels show the Fourier transform of the magnetic field and polarization along the direction $63^\circ$ relative to the $\hat{x}$ direction, denoted by the black dashed line in panel (a). Both the wavelength and polarization are similar to the waves observed in our 1D shock simulation and in the linear analysis, indicating that 1D shock simulations do capture  the  most  essential wave properties in the upstream of quasi-perpendicular shocks.

\begin{figure}
    \centering
    \includegraphics[width=0.4\textwidth]{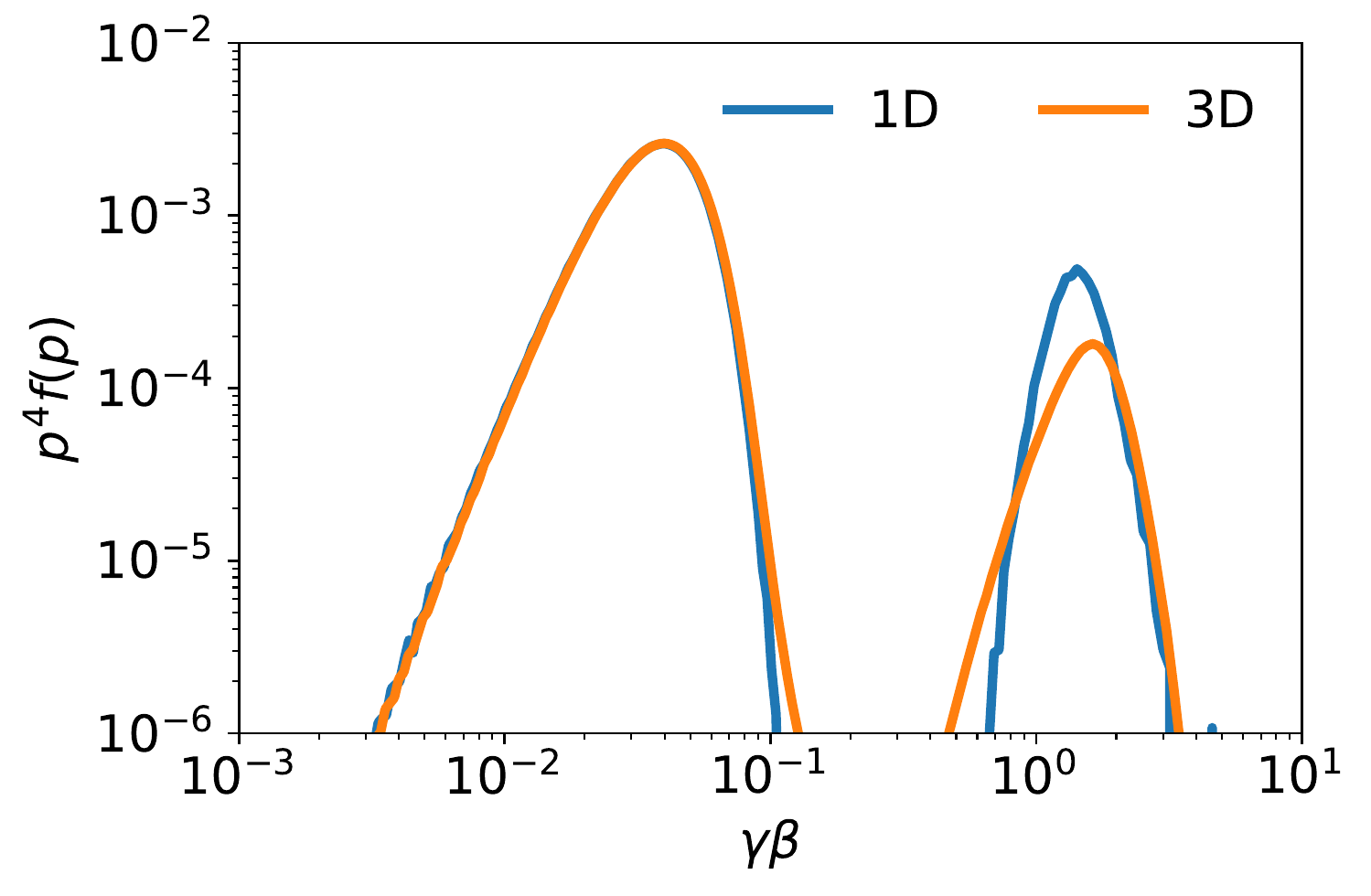}
    \caption{ Upstream electron spectra ($1-2 \times 10^3 c/\omega_{pe}$ relative to the shock front) at time $T\approx 3\Omega_{ci}$ for the 1D (blue line) and 3D (orange line) PIC shock simulations. Electron reflection efficiency in 3D simulations is lower by a factor of few compared to 1D simulations. For the 3D PIC simulation, we use transverse size $50c/\omega_{pe}$ in both $\hat{y}$ and $\hat{z}$ directions with 8 cells per electron skin depth. Other parameters are the same as in our 1D simulations. } 
    \label{fig:3d_spect}
\end{figure}

The results of 1D shock simulations should be further  verified in multi-dimensions. Typically, multi-dimensional simulations allow for larger variability of magnetic inclination at the shock (e.g., due to shock corrugation), and this can reduce the average efficiency of electron injection. 
To test this, we measured electron reflection in a short 3D shock simulation with the same parameters as our fiducial 1D runs (and transverse size of $50\times 50 (c/\omega_{pe})^2$).  
Figure \ref{fig:3d_spect} shows the comparison of upstream electron spectra for 1D and 3D PIC simulations at time $T\sim 3\Omega_{ci}^{-1}$. We see the electron reflection efficiency in 3D is lower but only by a factor of few compared to 1D simulations, which indicates that 1D simulations can capture the relevant range of reflectivities of 3D quasi-perpendicular shocks. We thus expect that the salient features of electron reflection, wave generation, and power law formation that we see in 1D will persist in future long-term multi-dimensional studies. 
The comparison with 1D simulations should be done with sufficiently large number of particles per cell to avoid artificial cooling of high-energy electrons due to discreteness effects in PIC simulations \citep{Kato2013}. In addition, most of our simulations were done with an artificial mass ratio of 100. We have done limited simulations at $m_i/m_e=400$, and find that the early properties of shocks, including the reflected electron fraction, are not sensitive to the mass ratio. However, higher mass ratio simulations require higher Alfv\`{e}nic Mach numbers in order to drive the upstream waves with electrons and reach the injection into DSA. These issues make a proper comparison with 1D simulations quite challenging numerically, and we plan to present this in an upcoming study.

We would like to thank Marian Lazar and Martin Weidl for useful discussions. This research was supported by NSF (grants AST-1814708, AST-1714658, AST-1909778, PHY-1804048, PHY-1748958, PHY-2010240) and by NASA (grants NNX17AG30G, 80NSSC18K1218 and 80NSSC18K1726) and by the International Space Science Institute’s (ISSI) International Teams program.
 The simulations in this paper were performed using computational resources at the TIGRESS high-performance computer center at Princeton University. AS is supported by Simons Foundation (grant 267233).

\bibliography{sh}{}
\bibliographystyle{apj}

\end{document}